NRF2011-EDU001-EL001 eduLab Project Scaling-up Reflections on Using Open Source Physics

Loo Kang Wee[1], Ai Phing Lim[2], Sze Yee Lye[1]

[1]Ministry of Education, Educational Technology Division (ETD), Singapore
[2]Ministry of Education, River Valley High School (RVHS), Singapore

lawrence_wee@moe.gov.sg, lim_ai_phing@moe.edu.sg, lye_sze_yee@moe.edu.sg

Abstract

eduLab (MOE, 2012b) is a key programme under the third MasterPlan (mp3) in Education harnessing information and communications technology (ICT) where teachers with good ideas for an ICT-enhanced lesson or curriculum (learning with computer models through inquiry, example PhET (PhET, 2011) can come together to collaborate. eduLab aims to support teachers to develop, prototype and test-bed their lesson ideas (journey in 2012-2014) while ensuring that the results, in the form of complete lesson packages (see http://edulab.moe.edu.sg/edulab-programmes/existing-projects  third project) , are scalable across schools to benefit the wider teaching community. Our models and lessons are downloadable here http://weelookang.blogspot.sg/2013/03/moe-excel-fest-2013-scaling.html. We have collaborated with namely Professor Francisco Esquembre, Fu-Kwun Hwang and Wolfgang Christian and created Open Source Computer Models on the topic of 1 Dimensional Collision (Loo Kang Wee, 2012b), Falling Magnet in Coil, Ripple Tank (Duffy, 2010; G. H. Goh et al., 2012; Ong, Ng, Goh, & Wee, 2012; Ong, Ng, Teo, et al., 2012; Loo Kang Wee, Duffy, Aguirregabiria, & Hwang, 2012), Superposition Waves, 2 Mass Gravity, Earth-Moon Gravity, Kepler's Solar System and Geostationary Orbit (J. Goh & Wee, 2011; Loo Kang Wee, 2012a; Loo Kang Wee & Esquembre, 2010; Loo Kang Wee & Goh, 2013) for enriching interactive engagement (Christian & Belloni, 2000; Hake, 1998) previously lacking in our teaching practices, but we argue is essential for deepening learning (L.K. Wee & Lye, 2012) by doing. We hope to share some of the scaling-up learning points from ideas (researchers Francisco Fu-Kwun and Wolfgang) to practice (teachers in RVHS, YJC, IJC, AJC and SRJC) supported by specialist (ETD), contextualized and made possible through this edulab funding initiative by National Research Fund, managed by National Institute of Education (NIE) and Ministry of Education, Singapore.

Scaling–Up Reflections on Using Open Source Physics

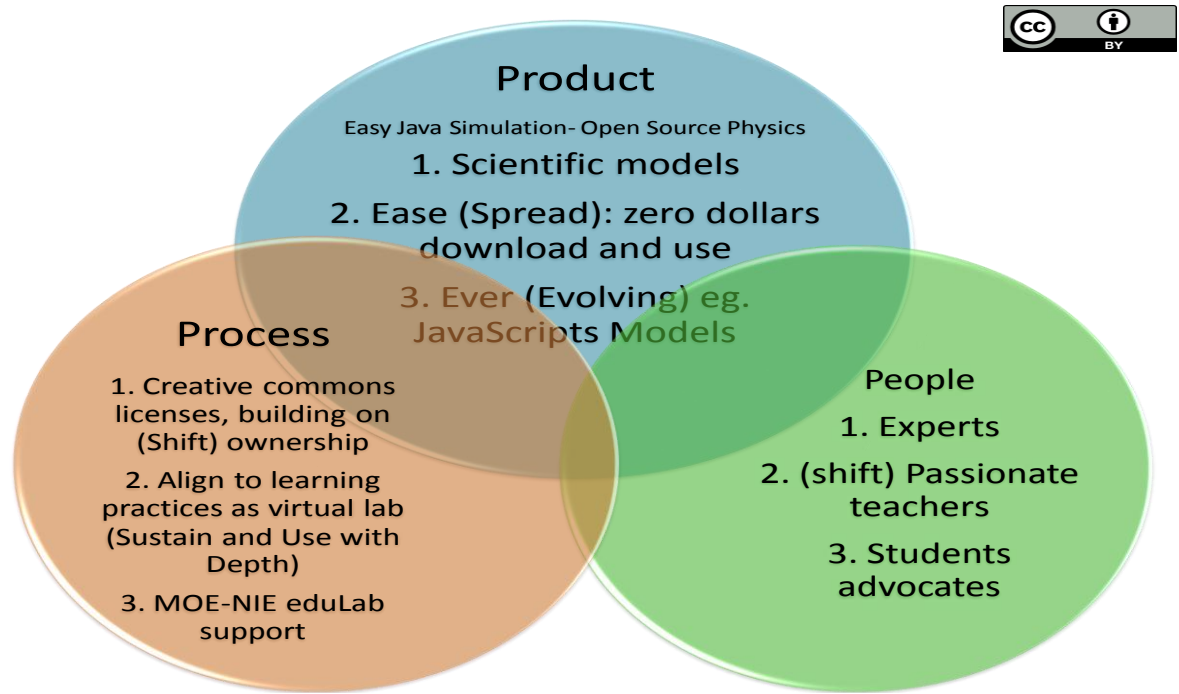

Figure 1. 3Ps Scaling–Up Framework, Product, Process and People with Dimensions of Scaling Up in brackets (Dede, 2007).

Product[3] – Problem – Purpose:

When the product[1] Easy Java Simulations (Esquembre, 2012; Loo Kang Wee, Goh, & Lim, 2014) (EJS) authoring tool–kit and Open Source Physics (Christian & Esquembre, 2012; Christian & Titus, 1998) (OSP) (Physlets) came together, Professor Francisco Esquembre and Wolfgang Christian made the design, creation and redistribution of scientific-mathematical computer models (Loo Kang Wee, Lim, et al., 2012), even easier than before.

With the problem of Physics learning in the absence of any hand-on activities, simulations could be a possible tool to supplement experiential learning or compliment real equipment (Loo K. Wee & Ning, 2014) laboratory experiments. Teachers usually rely on anything such as video, animations or simulations usually not editable by the teachers' themselves. EJS does a great job to produce[2] free (download and use), accurate and scientific models quickly that can be studied (open source codes), remixed and reused. There is also no need for login or server setup (Vargas et al., 2008), requiring only Java runtime (for *.jar EJS models) or a modern browser ( EJSS JavaScripts models).

As of September 2013 during 18$^{th}$ Multimedia Physics Teaching and Learning conference, Madrid, Spain, Francisco released EJS5.0 capable of producing[3]-generating JavaScripts computer models, viewable on almost any mobile operating system like Android and iOS, simply made the already great product, even better still to achieve the purpose of providing quality open educational interactive resources.

Process[3] – Practice:

Licensed creative commons attribution (L.K. Wee & Lye, 2012) or other compatible licenses, practically only just requiring an internet connection to download the computer models with the source codes included, is a real game changer for scaling–up use of computer models. In addition, permission to legally customize these computer models is made explicitly clear is one of the key processes[1] we identified, supported with OSP discussion forum help experts and others all over the world.

Aligning to existing practice[2] of laboratory (Baser & Durmus, 2010; Dormido et al., 2008; Espinoza & Quarless, 2010; Jara et al., 2009), many of the EJS models can be used as a virtual laboratory to support experiential learning (Loo Kang Wee, 2012b). Since science educational practices regularly requires students to conduct hands-on experiments to inquiry about the physics phenomena, it is not surprising as students and teachers are more comfortable using them to support their learning as oppose to game playing (Chee, Tan, Tan, & Jan, 2011; Jan, Chee, & Tan, 2010; Squire, 2006) as a form of science learning.

We also acknowledge that it is useful to acquire funding[3] support especially in allowing experts to share their knowledge in person and supporting teachers to spend time on this project.

People[3] – Passion

The people[1] – experts like Professor Francisco Esquembre, Fu-Kwun Hwang and Wolfgang Christian and their communities in the world really shown the thought leadership of what it means to scaling–up meaningful use of technology, as explained using the 3P and Dimensions of Scaling–up.

Now ordinary but passionate educator(s)[2] can now add on or create finer customized computer models to suit their technology, pedagogy, content and context knowledge to better mould the learning experiences of their students. In any learning community, the key people keep creating more computer models to suit their fellow teachers and students learning needs and re-released these computer models with activity worksheets and other resources for the benefit of all, are the real motivation that drives our collaborative work using EJS.

*Area of improvement*

As for an area of improvement, we find that having more students[3] advocates to be an useful indicator of scaling–up as evident in projects like Scratch (Resnick et al., 2009) and GeoGebra (Hohenwarter & Fuchs, 2004) to be useful projects for which EJS can learn from, in terms of scaling–up.

Inspire through giving back: Our Models and Curriculum Developed

Our Project artifacts are licensed creative commons attribution and downloadable from http://edulab.moe.edu.sg/edulab-programmes/existing-projects such as the 6th Advance level physics instructional program support group (6[th] IPSG A-Level Physics) link and MOE Excellence through Continuous Enterprise & Learning (ExCEL Fest) link.

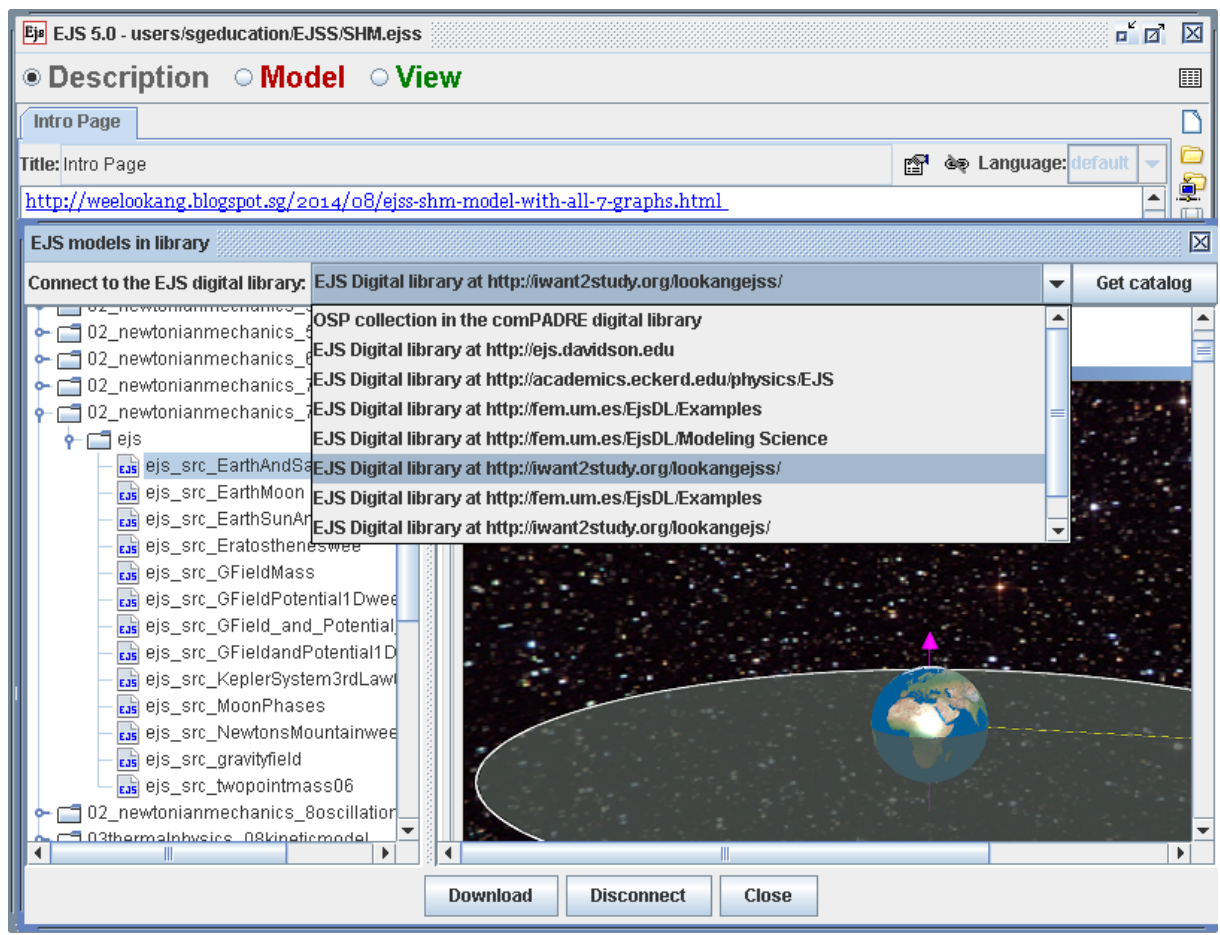

Figure 2.   EJS 5.0 authoring toolkit view of the Read from Digital Library and select the EJS Digital library at http://iwant2study.org/lookangejss/

To inspire fellow educators to look into open educational resources and enhance the scaling–up of our models, it is now possible to directly access through EJS authoring toolkit under Shared Library, downloadable and editable through the creation of the Singapore digital library via http://iwant2study.org/lookangejss/.

Acknowledgement

We wish to acknowledge the passionate contributions of Francisco Esquembre, Fu-Kwun Hwang, Wolfgang Christian, Andrew Duffy, Todd Timberlake and Juan

Aguirregabiria and many more in the OSP community for their ideas and insights in the co-creation of interactive simulation and curriculum materials.

This research is made possible thanks to the eduLab project NRF2011-EDU001-EL001 Java Simulation Design for Teaching and Learning, awarded by the National Research Foundation (NRF), Singapore in collaboration with National Institute of Education (NIE), Singapore and the Ministry of Education (MOE), Singapore.

We also thank MOE for the recognition of our research on computer model lessons as a significant innovation in 2012 MOE Innergy (HQ) GOLD (MOE, 2012a) and commendation 2014 Awards by Educational Technology Division and Academy of Singapore Teachers.

Any opinions, findings, conclusions or recommendations expressed in this paper, are those of the authors and do not necessarily reflect the views of the MOE, NIE or NRF.